\documentstyle[12pt]{article}
\def\ie{{\em i.e.}}
\def\eg{{\em e.g.}}

\def\Zbb{Z\rightarrow b\ov b}

\def\epsb{\epsilon_b}
\def\tanb{\tan\beta}
\def\beq{\begin{equation}}
\def\eeq{\end{equation}}

\catcode`\@=11 

\def\lsim{\mathrel{\mathpalette\@versim<}}
\def\gsim{\mathrel{\mathpalette\@versim>}}
\def\@versim#1#2{\vcenter{\offinterlineskip
    \ialign{$\m@th#1\hfil##\hfil$\crcr#2\crcr\sim\crcr } }}
\def\etal{{\em et. al.}}
\def\JL{J. L. Lopez}
\def\DVN{D. V. Nanopoulos}

\def\r#1{$\bf#1$}
\def\rb#1{$\bf\overline{#1}$}

\def\t1{{\tilde 1}}
\def\ov{\overline}

\def\GeV{\,{\rm GeV}}
\def\TeV{\,{\rm TeV}}

\def\to{\rightarrow}

\def\NPB#1#2#3{Nucl. Phys. B {\bf#1} (19#2) #3}
\def\PLB#1#2#3{Phys. Lett. B {\bf#1} (19#2) #3}

\def\PRD#1#2#3{Phys. Rev. D {\bf#1} (19#2) #3}
\def\PRL#1#2#3{Phys. Rev. Lett. {\bf#1} (19#2) #3}

\def\TAMU#1{Texas A \& M University preprint CTP-TAMU-#1}

\begin{document}
\begin{flushright}
\baselineskip=12pt
{CERN-TH.7078/93}\\
{CTP-TAMU-68/93}\\
{ACT-24/93}\\
\end{flushright}
%

\begin{center}
\vglue 0.3cm
{\Large\bf New Precision Electroweak Tests of\\}
\vspace{0.2cm}
{\Large\bf SU(5) x U(1) Supergravity\\}
\vglue 0.5cm
{JORGE L. LOPEZ$^{(a),(b)}$, D. V. NANOPOULOS$^{(a),(b),(c)}$,
GYE T. PARK$^{(a),(b)}$,\\}
{and A. ZICHICHI$^{(d)}$\\}
\vglue 0.4cm
{\em $^{(a)}$Center for Theoretical Physics, Department of Physics, Texas A\&M
University\\}
{\em College Station, TX 77843--4242, USA\\}
{\em $^{(b)}$Astroparticle Physics Group, Houston Advanced Research Center
(HARC)\\}
{\em Mitchel Campus, The Woodlands, TX 77381, USA\\}
{\em $^{(c)}$CERN, Theory Division, 1211 Geneva 23, Switzerland\\}
{\em $^{(d)}$CERN, 1211 Geneva 23, Switzerland\\}
\baselineskip=12pt

\vglue 0.4cm
{\tenrm ABSTRACT}
\end{center}
\vglue -0.2cm
{\rightskip=3pc
 \leftskip=3pc
\noindent
We explore the one-loop electroweak radiative corrections in $SU(5)\times U(1)$
supergravity via explicit calculation of vacuum-polarization and
vertex-correction contributions to the $\epsilon_1$ and $\epsilon_b$
parameters. Experimentally, these parameters are obtained from a
global fit to the set of observables $\Gamma_{l}, \Gamma_{b}, A^{l}_{FB}$,
and $M_W/M_Z$. We include $q^2$-dependent effects, which induce a large
systematic negative shift on $\epsilon_{1}$ for light chargino masses
($m_{\chi^\pm_1}\lsim70\GeV$). The (non-oblique) supersymmetric vertex
corrections to $\Zbb$, which define the $\epsilon_b$ parameter, show
a significant positive shift for light chargino masses, which for
$\tan\beta\approx2$ can be nearly compensated by a negative shift from the
charged Higgs contribution.  We conclude that at the 90\%CL, for
$m_t\lsim160\GeV$ the present experimental values of $\epsilon_1$ and
$\epsilon_b$ do not constrain in any way $SU(5)\times U(1)$ supergravity in
both no-scale and dilaton scenarios. On the other hand, for $m_t\gsim160\GeV$
the constraints on the parameter space become increasingly stricter. We
demonstrate this trend with a study of the $m_t=170\GeV$ case, where only a
small region of parameter space, with $\tan\beta\gsim4$, remains allowed and
corresponds to light chargino masses ($m_{\chi^\pm_1}\lsim70\GeV$). Thus
$SU(5)\times U(1)$ supergravity combined with high-precision LEP data would
suggest the presence of light charginos if the top quark is not detected at the
Tevatron.}
\vspace{1cm}
\begin{flushleft}
\baselineskip=12pt
{CERN-TH.7078/93}\\
{CTP-TAMU-68/93}\\
{ACT-24/93}\\
October 1993
\end{flushleft}
\vfill\eject
\setcounter{page}{1}
\pagestyle{plain}

\baselineskip=14pt

\section{Introduction}
Since the advent of LEP, precision electroweak tests have become rather deep
probes of the Standard Model of electroweak interactions and its challengers.
These tests have demonstrated the internal consistency of the Standard Model,
as long as the yet-to-be-measured top-quark mass ($m_t$) is within certain
limits, which depend on the value assumed for the Higgs-boson mass ($m_H$):
$m_t=135\pm18\GeV$ for $m_H\sim60\GeV$ and $m_t=174\pm15\GeV$ for
$m_H\sim1\TeV$ (for a recent review see \eg, Ref.~\cite{Altlecture}). In the
context of supersymmetry, such tests have been performed throughout the
years within the Minimal Supersymmetric Standard Model (MSSM)
\cite{OldEW,BFC,ABC,ABCII}. The problem with such calculations is well known
but usually ignored -- there are too many parameters in the MSSM (at least
twenty) -- and therefore it is not possible to obtain precise predictions for
the observables of interest.

In the context of supergravity models, on the other hand, any observable can be
computed in terms of at most five parameters: the top-quark mass, the ratio of
Higgs vacuum expectation values ($\tan\beta$), and three universal
soft-supersymmetry-breaking parameters $(m_{1/2},m_0,A)$ \cite{Erice93}. This
implies much sharper predictions for the various quantities of interest, as
well as numerous correlations among them. Of even more experimental interest is
$SU(5)\times U(1)$ supergravity where string-inspired ans\"atze for the
soft-supersymmetry-breaking parameters allow the theory to be described in
terms of only three parameters: $m_t$, $\tan\beta$, and $m_{\tilde g}$
\cite{EriceDec92}. Precision electroweak tests in the no-scale
\cite{LNZI} and dilaton \cite{LNZII} scenarios for $SU(5)\times U(1)$
supergravity have been performed in Refs.~\cite{ewcorr,bsg-eps}, using the
description in terms of the $\epsilon_{1,2,3}$ parameters introduced in
Refs.~\cite{AB,ABJ}. In this paper we extend these tests in two ways: first,
we include for the first time the $\epsilon_b$ parameter \cite{ABC} which
encodes the one-loop corrections to the $Z\to b\bar b$ vertex, and second
we perform the calculation of the $\epsilon_1$ parameter in a new scheme
\cite{ABC}, which takes full advantage of the latest experimental data.

The calculation of $\epsilon_b$ is of particular importance since in the
Standard Model, of the four parameters $\epsilon_{1,2,3,b}$ at present only
$\epsilon_b$ falls outside the 1$\sigma$ experimental error (for $m_t>120\GeV$)
\cite{ABC,BV}. This discrepancy is not of great statistical
significance, although the trend should not be overlooked, especially in the
light of the much better statistical agreement for the other three parameters.
Within the context of the Standard Model, another reason for focusing attention
on the $\epsilon_{b}$ parameter is that, unlike the $\epsilon_1$ parameter,
$\epsilon_b$ provides a constraint on the top-quark mass which is {\em
practically independent} of the Higgs-boson mass. Indeed, at the 95\% CL,
the limits on $\epsilon_b$ require $m_t<185\GeV$, whereas those from
$\epsilon_1$ require $m_t<177-198\GeV$ for $m_H\sim100-1000\GeV$ \cite{BV}.

In supersymmetric models, the weakening of the $\epsilon_1$-deduced $m_t$ upper
bound for large Higgs-boson masses does not occur (since the Higgs boson must
be light) and both $\epsilon_1$ and $\epsilon_b$ are expected to yield
comparable constraints. In this context it has been pointed out \cite{ABCII}
that if certain mass correlations in the MSSM are satisfied, then the
prediction for $\epsilon_b$ will be in better agreement with the data than the
Standard Model prediction is. However, the opposite situation could also occur
(\ie, worse agreement), as well as negligble change relative to the Standard
Model prediction (when all supersymmetric particles are heavy enough). We show
that this three-way ambiguity in the MSSM prediction for $\epsilon_b$
disappears when one considers $SU(5)\times U(1)$ supergravity in both no-scale
and dilaton scenarios. The $SU(5)\times U(1)$ supergravity prediction is
practically always in better statistical agreement with the data (compared with
the Standard Model one).

This study shows that at the 90\%CL, for $m_t\lsim160\GeV$ the present
experimental values of $\epsilon_1$ and $\epsilon_b$ do not constrain
$SU(5)\times U(1)$ supergravity in any way. On the other hand, for
$m_t\gsim160\GeV$ the constraints on the parameter space become increasingly
stricter.  We demonstrate this trend with a study of the $m_t=170\GeV$ case,
where only a small region of parameter space, with $\tan\beta\gsim4$, remains
allowed and corresponds to a light supersymmetric spectrum, and in
particular light chargino masses ($m_{\chi^\pm_1}\lsim70\GeV$). Thus
$SU(5)\times U(1)$ supergravity combined with high-precision LEP data would
suggest the presence of light charginos if the top quark is not detected at the
Tevatron.

\section{SU(5)xU(1) Supergravity}
Our study of one-loop electroweak radiative corrections is performed within
the context of $SU(5)\times U(1)$ supergravity \cite{EriceDec92}. Besides
the several theoretical string-inspired motivations that underlie this theory,
of great practical importance is the fact that only three parameters are needed
to describe all their possible predictions. This fact has been used in the
recent past to perform a series of calculations for collider
\cite{collider,LNPWZh} and rare \cite{rare,ewcorr,bsg-eps} processes within
this theory. The constraints obtained from all these analyses should help
sharpen even more the experimental predictions for the remaining allowed points
in parameter space.

In $SU(5)\times U(1)$ supergravity, gauge coupling unification occurs at the
string scale $10^{18}\GeV$ \cite{EriceDec92}, because of the presence of a pair
of \r{10},\rb{10} representations with intermediate-scale masses. The three
parameters alluded to above are: (i) the top-quark mass ($m_t$), (ii) the ratio
of Higgs vacuum expectation values ($\tan\beta$), which satisfies
$1\lsim\tan\beta\lsim40$, and (iii) the gluino mass, which is cut off at 1 TeV.
This simplification in the number of input parameters is possible because of
specific string-inspired scenarios for the universal
soft-supersymmetry-breaking parameters ($m_0,m_{1/2},A$) at the unification
scale. These three parameters can be computed in specific string models in
terms of just one of them \cite{IL}. In the {\em no-scale} scenario one obtains
$m_0=A=0$, whereas in the {\em dilaton} scenario the result is
$m_0=\frac{1}{\sqrt{3}}m_{1/2}, A=-m_{1/2}$.
After running the renormalization group equations from high to low energies,
at the low-energy scale the requirement of radiative electroweak symmetry
breaking introduces two further constraints which determine
the magnitude of the Higgs mixing term $\mu$, although its sign remains
undetermined. Finally, all the known phenomenological constraints on the
sparticle masses are imposed (most importantly the chargino, slepton, and Higgs
mass bounds). This procedure is well documented in the literature
\cite{aspects} and yields the allowed parameter spaces for the no-scale
\cite{LNZI} and dilaton \cite{LNZII} scenarios.

These allowed parameter spaces in the three defining variables
($m_t,\tan\beta,m_{\tilde g}$) consist of a discrete set of points for
three values of $m_t$ ($m_t=130,150,170\GeV$), and a discrete set of allowed
values for $\tan\beta$, starting at 2 and running (in steps of two) up to 32
(46) for the no-scale (dilaton) scenario. The chosen lower bound on $\tan\beta$
follows from the requirement by the radiative breaking mechanism of
$\tan\beta>1$, and because the LEP lower bound on the lightest Higgs boson mass
($m_h\gsim60\GeV$ \cite{LNPWZh}) is quite constraining for $1<\tan\beta<2$.

\begin{table}
\hrule
\caption{The approximate proportionality coefficients to the gluino mass, for
the various sparticle masses in the two supersymmetry breaking scenarios
considered.}
\label{Table1}
\begin{center}
\begin{tabular}{|c|c|c|}\hline
&no-scale&dilaton\\ \hline
$\tilde e_R,\tilde \mu_R$&$0.18$&$0.33$\\
$\tilde\nu$&$0.18-0.30$&$0.33-0.41$\\
$2\chi^0_1,\chi^0_2,\chi^\pm_1$&$0.28$&$0.28$\\
$\tilde e_L,\tilde \mu_L$&$0.30$&$0.41$\\
$\tilde q$&$0.97$&$1.01$\\
$\tilde g$&$1.00$&$1.00$\\ \hline
\end{tabular}
\end{center}
\hrule
\end{table}

In the models we consider all sparticle masses scale with the gluino mass, with
a mild $\tan\beta$ dependence (except for the third-generation squark and
slepton masses). In Table~\ref{Table1} we give the approximate
proportionality coefficient (to the gluino mass) for each sparticle mass. Note
that the relation $2m_{\chi^0_1}\approx m_{\chi^0_2}\approx m_{\chi^\pm_1}$
holds to good approximation. The third-generation squark and slepton masses
also scale with $m_{\tilde g}$, but the relationships are smeared by a strong
$\tan\beta$ dependence. From Table~\ref{Table1} one can (approximately)
translate any bounds on a given sparticle mass on bounds on all the other
sparticle masses.

\section{One-loop electroweak radiative corrections and the new $\epsilon$
parameters}
There are different schemes to parametrize the electroweak (EW) vacuum
polarization corrections \cite{Kennedy,PT,efflagr,AB}. It can be shown, by
expanding the vacuum polarization tensors to order $q^2$, that one obtains
three independent physical parameters. Alternatively, one can show that upon
symmetry breaking three additional terms appear in the effective lagrangian
\cite{efflagr}. In the $(S,T,U)$ scheme \cite{PT}, the deviations of the model
predictions from the SM predictions (with fixed SM values for $m_t,m_{H_{SM}}$)
are considered as the effects from ``new physics". This scheme is only valid to
the lowest order in $q^2$, and is therefore not applicable to a theory with
new, light $(\sim M_Z)$ particles. In the $\epsilon$-scheme\cite{ABJ,ABC}, on
the other hand, the model predictions are absolute and also valid up to higher
orders in $q^2$, and therefore this scheme is more applicable to the EW
precision tests of the MSSM \cite{BFC} and a class of supergravity models
\cite{ewcorr}.

There are two different $\epsilon$-schemes. The original scheme\cite{ABJ} was
considered in our previous analyses \cite{ewcorr,bsg-eps}, where
$\epsilon_{1,2,3}$ are defined from a basic set of observables $\Gamma_{l},
A^{l}_{FB}$ and $M_W/M_Z$.
Due to the large $m_t$-dependent vertex corrections to $\Gamma_b$, the
$\epsilon_{1,2,3}$ parameters   and $\Gamma_b$ can be correlated only for a
fixed value of $m_t$. Therefore, $\Gamma_{tot}$, $\Gamma_{hadron}$ and
$\Gamma_b$ were not included  in Ref.~\cite{ABJ}. However, in the new
$\epsilon$-scheme, introduced recently in Ref.~\cite{ABC}, the above
difficulties are overcome by introducing a new parameter $\epsilon_b$ to encode
the $\Zbb$ vertex corrections. The four $\epsilon$'s are now defined from an
enlarged set of $\Gamma_{l}$, $\Gamma_{b}$, $A^{l}_{FB}$ and $M_W/M_Z$ without
even specifying $m_t$.
In this work we use this new $\epsilon$-scheme.
Experimentally, including all LEP data allows one to
determine the allowed ranges for these parameters \cite{Altlecture}
\beq
\epsilon^{exp}_1=(-0.3\pm3.2)\times10^{-3},\qquad
\epsilon^{exp}_b=(3.1\pm5.5)\times10^{-3}\ .
\eeq
Since among $\epsilon_{1,2,3}$ only $\epsilon_1$ provides constraints in
supersymmetric models at the 90\%CL \cite{ewcorr,ABCII}, we discuss below only
$\epsilon_1$ and $\epsilon_b$.

The expression for $\epsilon_1$ is given as
\cite{BFC}
\beq
\epsilon_1=e_1-e_5-{\delta G_{V,B}\over G}-4\delta g_A,\label{eps1}
\eeq
where $e_{1,5}$ are the following combinations of vacuum polarization
amplitudes
\begin{eqnarray}
e_1&=&{\alpha\over 4\pi \sin^2\theta_W M^2_W}[\Pi^{33}_T(0)-\Pi^{11}_T(0)],
\label{e1}\\
e_5&=& M_Z^2F^\prime_{ZZ}(M_Z^2),\label{e5}
\end{eqnarray}
and the $q^2\not=0$ contributions $F_{ij}(q^2)$ are defined by
\beq
\Pi^{ij}_T(q^2)=\Pi^{ij}_T(0)+q^2F_{ij}(q^2).
\eeq
The $\delta g_A$ in Eqn.~(\ref{eps1}) is the contribution to the axial-vector
form factor at $q^2=M^2_Z$ in the $Z\to l^+l^-$ vertex from proper vertex
diagrams and fermion self-energies, and $\delta G_{V,B}$ comes from the
one-loop box, vertex and fermion self-energy corrections to the $\mu$-decay
amplitude at zero external momentum. These non-oblique SM corrections are
non-negligible, and must be included in order to obtain an accurate SM
prediction.
As is well known, the SM contribution to $\epsilon_1$ depends quadratically
on $m_t$ but only logarithmically on the SM Higgs boson mass ($m_H$). In this
fashion upper bounds on $m_t$ can be obtained which have a non-negligible $m_H$
dependence: up to $20\GeV$ stronger when going from a heavy ($\approx1\TeV$)
to a light ($\approx100\GeV$) Higgs boson. It is also known (in the MSSM) that
the largest supersymmetric contributions to $\epsilon_1$ are expected to
arise from the $\tilde t$-$\tilde b$ sector, and in the limiting case of a very
light stop, the contribution is comparable to that of the $t$-$b$ sector. The
remaining squark, slepton, chargino, neutralino, and Higgs sectors all
typically contribute considerably less. For increasing sparticle masses, the
heavy sector of the theory decouples, and only SM effects  with a {\it light}
Higgs boson survive. (This entails stricter upper bounds on $m_t$ than in the
SM, since there the Higgs boson does not need to be light.) However, for a
light chargino ($m_{\chi^\pm_1}\to{1\over2}M_Z$), a $Z$-wavefunction
renormalization threshold effect can introduce a substantial $q^2$-dependence
in the calculation, \ie, the presence of $e_5$ in Eq.~(\ref{eps1}) \cite{BFC}.
The complete vacuum polarization contributions from the Higgs sector, the
supersymmetric chargino-neutralino and sfermion sectors, and also the
corresponding contributions in the SM have been included in our calculations
\cite{ewcorr}.

Following Ref.~\cite{ABC}, $\epsb$ is defined from $\Gamma_b$, the inclusive
partial width for $\Zbb$, as follows

\begin{equation}
\Gamma_b=3 R_{QCD} {G_FM^3_Z\over 6\pi\sqrt 2}\left(
1+{\alpha\over 12\pi}\right)\left[ \beta _b{\left( 3-\beta
^2_b\right)\over 2}(g^b_V)^2+\beta^3_b (g^b_A)^2\right] \;,
\end{equation}
with
\begin{eqnarray}
R_{QCD} &\cong&\left[1+1.2{\alpha_S\left(
M_Z\right)\over\pi}-1.1{\left(\alpha_S\left(
M_Z\right)\over\pi\right)}^2-12.8{\left(\alpha_S\left(
M_Z\right)\over\pi\right)}^3\right] \;,\\
\beta_b&=&\sqrt {1-{4m_b^2\over M_Z^2}} \;, \\
g^b_A&=&-{1\over2}\left(1+{\epsilon_1\over2}\right)\left(
1+{\epsb}\right)\;,\\
{g^b_V\over{g^b_A}}&=&{{1-{4\over3}{\ov s}^2_W+\epsb}\over{1+\epsb}}\;.
\end{eqnarray}
Here ${\ov s}^2_W$ is an effective $\sin^2\theta_W$ for on-shell $Z$, and
$\epsb$ is closely related to the real part of the vertex correction to $\Zbb$,
denoted in the literature by $\nabla_b$ and defined explicitly in
Ref.~\cite{BF}.  In the SM, the diagrams for $\nabla_b$  involve top quarks and
$W^\pm$ bosons \cite{RbSM}, and the contribution to $\epsb$ depends
quadratically on $m_t$. In supersymmetric models there are additional diagrams
involving Higgs bosons and supersymmetric particles. The charged Higgs
contributions have been calculated in Refs.~\cite{Denner,Rbbsg2HD,epsb2HD} in
the context of a non-supersymmetric two Higgs doublet model, and the
contributions involving supersymmetric particles in Refs.~\cite{BF,Rb2HD}.
Moreover, $\epsilon_b$ itself has been calculated in Ref.~\cite{epsb2HD}.
The additional supersymmetric contributions are: (i) a negative contribution
from charged Higgs--top exchange which grows as $m^2_t/\tan^{2}\beta$ for
$\tan\beta\ll{m_t\over{m_b}}$; (ii) a positive contribution from chargino-stop
exchange which in this case grows as $m^2_t/\sin^{2}\beta$; and (iii) a
contribution from neutralino(neutral Higgs)--bottom exchange which grows as
$m^2_b\tan^{2}\beta$ and is negligible except for large values of $\tan\beta$
(\ie, $\tan\beta\gsim{m_t\over{m_b}}$) (the contribution (iii) has been
neglected in our analysis).

\section{Results and discussion}
In Figures 1--4 we show the results of the calculation of $\epsilon_1$ and
$\epsilon_b$ (as described above) for all the allowed points in $SU(5)\times
U(1)$ supergravity in both no-scale and dilaton scenarios. Since all sparticle
masses nearly scale with the gluino mass (or the chargino mass), it suffices to
show the dependences of these parameters on, for example, the chargino mass.
Table 1 can be used to deduce the dependences on any of the other masses. We
only show the explicit dependence on the chargino mass (in Figs.~1,3) for the
case $m_t=170\GeV$, since for $m_t=130,150\GeV$ there are no constraints at the
90\%CL. However, in the correlated $(\epsilon_1,\epsilon_b)$ plots (Figs.~2,4)
we show the results for all three values of $m_t$.

The qualitative results for $\epsilon_1$ are similar to those obtained in
Refs.~\cite{ewcorr,bsg-eps} using the old definition of $\epsilon_1$. That is,
for light chargino masses there is a large negative shift due to a threshold
effect in the $Z$-wavefunction renormalization for $m_{\chi^\pm_1}\to
{1\over2}M_Z$ (as first noticed in Ref.~\cite{BFC}). As soon as the sparticle
masses exceed $\sim100\GeV$ the result quickly asymptotes to the Standard Model
value for a light Higgs boson mass ($\lsim100\GeV$). Quantitatively, the
enlarged set of observables in the new $\epsilon$-scheme shifts the
experimentally allowed range somewhat and the bounds become slightly weaker
than in Refs.~\cite{ewcorr,bsg-eps}. These remarks apply to both no-scale and
dilaton scenarios.

In the case of $\epsilon_b$, the results also asymptote to the Standard Model
values for large sparticle masses as they should. Two competing effects are
seen to occur: (i) a positive shift for light chargino masses, and (ii) and
negative shift for light charged Higgs masses and small values of $\tan\beta$.
In fact, the latter effect becomes evident in Figures 1,3 (bottom rows) as the
solid curve corresponding to $\tan\beta=2$. What happens here is that the
charged Higgs contribution nearly cancels the chargino contribution \cite{BF},
making $\epsilon_b$ asymptote much faster to the SM value.

We also notice from Figure 3 (bottom row) that there are lines of points far
below the solid curve corresponding to $\tan\beta=2$ in the dilaton scenario.
These correspond to {\em large} $\tan\beta(\gsim {m_t\over{m_b}})$ for which
the charged Higgs diagram gets a significant contribution $\sim
m^2_b\tan^{2}\beta$ coming from the charged Higgs coupling to $b_R$. Such large
values of $\tan\beta$ are not allowed in the no-scale scenario. It must be
emphasized that for such large values of $\tan\beta$, the neglected
neutralino--neutral Higgs diagrams will also become significant \cite{BF} and
since especially neutralino diagrams give a positive contribution, their effect
could compensate the large negative charged Higgs contributions.

For $m_t=170\GeV$ at the 90\%CL one can safely exclude values of
$\tan\beta\lsim2$ in the no-scale and dilaton (except for just one point for
$\mu<0$) scenarios. Moreover, as Figs.~1,3 show, there are excluded points for
all values of $\tan\beta$. In the dilaton scenario, large values of $\tan\beta$
(\ie, $\tanb\gsim32$ for $\mu>0$ and $\tanb\gsim24$ for $\mu<0$) are also
constrained, and even perhaps excluded if the neutralino--neutral-Higgs
contributions are not large enough to compensate for these values.

It is seen that for light chargino masses and not too small values of
$\tan\beta$, the fit to the $\epsilon_b$ data is better in $SU(5)\times
U(1)$ supergravity than in the  Standard Model, although only marginally so.
To see the combined effect of $\epsilon_{1,b}$ for increasing values of $m_t$,
in Figs.~2,4 we show the calculated values of these parameters for
$m_t=130,150,170\GeV$, as well as the $1\sigma$ experimental ellipse (from
Ref.~\cite{ABCII}). Clearly smaller values of $m_t$ fit the data better.

\section{Conclusions}

We have computed the one-loop electroweak corrections in the form of the
$\epsilon_1$ and $\epsilon_b$ parameters in the context of $SU(5)\times U(1)$
supergravity in both no-scale and dilaton scenarios. The new $\epsilon$-scheme
used allows to include in the experimental constraints all of the LEP data. In
addition, the minimality of parameters in $SU(5)\times U(1)$ supergravity is
such that rather precise predictions can be made for these observables and this
entails strict constraints on the parameter spaces of the two scenarios
considered.

In agreement with our previous analysis, we find that for $m_t\lsim160\GeV$, at
the 90\%CL these constraints are not restricting at present. However, their
quadratic dependence on $m_t$ makes them quite severe for increasingly large
values of $m_t$. We have studied explicitly the case of $m_t=170\GeV$ and shown
that most points in parameter space are excluded. The exceptions occur for
light chargino masses which shift $\epsilon_1$ down and $\epsilon_b$ up.
However, for $\tan\beta\lsim2$ the $\epsilon_b$ constraint is so strong that no
points are allowed in the no-scale scenario.

In the near future, improved experimental sensitivity on the $\epsilon_b$
parameter is likely to be a decisive test of $SU(5)\times U(1)$ supergravity.
In any rate, the trend is clear: lighter values of the top-quark mass fit the
data much better than heavier ones do. In addition, supesymmetry seems to
always help in this statistical agreement. Finally, if the top quark continues
to remain undetected at the Tevatron, high-precision LEP data in the context
of $SU(5)\times U(1)$ supergravity would suggest the presence of light
charginos.

\section*{Acknowledgements}
This work has been supported in part by DOE grant DE-FG05-91-ER-40633. The work
of G.P. has been supported by a World Laboratory Fellowship.
G.P. thanks Michael Boulware and Donnald Finnell for very helpful discussions.

\newpage

\noindent{\large\bf Figure Captions}
\begin{description}
\item Figure 1: The predictions for the $\epsilon_1$ (top row) and $\epsilon_b$
(bottom row) parameters versus the chargino mass in the no-scale $SU(5)\times
U(1)$ supergravity scenario for $m_t=170\GeV$. In the top (bottom) row, points
between (above) the horizontal line(s) are allowed at the 90\% CL. The solid
curve (bottom row) represents the $\tanb=2$ line.
\item Figure 2: The correlated predictions for the $\epsilon_1$ and
$\epsilon_b$ parameters in $10^{-3}$ in the no-scale $SU(5)\times U(1)$
supergravity scenario. The ellipse represents the $1\sigma$ contour obtained
from all LEP data. The values of $m_t$ are as indicated.
\item Figure 3: The predictions for the $\epsilon_1$ (top row) and $\epsilon_b$
 (bottom row) parameters versus the chargino mass in dilaton $SU(5)\times U(1)$
supergravity scenario for $m_t=170\GeV$. In the top (bottom) row, points
between (above) the horizontal line(s) are allowed at the 90\% CL. The solid
curve (bottom row) represents the $\tanb=2$ line.
\item Figure 4: The correlated predictions for the $\epsilon_1$ and
$\epsilon_b$ parameters in $10^{-3}$ in the dilaton $SU(5)\times U(1)$
supergravity scenario. The ellipse represents the $1\sigma$ contour obtained
from all LEP data. The values of $m_t$ are as indicated.
\end{description}

\end{document}